\documentclass[journal=ancac3,manuscript=article]{achemso}

\def\ii{{\rm i}}  \def\ee{{\rm e}}    
      
      \def\rb{{\bf r}}

\usepackage[T1]{fontenc}
\usepackage[latin1]{inputenc}
\usepackage{lmodern}\usepackage{graphicx}
\usepackage{dcolumn}
\usepackage{bm}
\usepackage{amsmath}
\usepackage{amssymb}
\usepackage{pst-all}
\usepackage{psfrag}
\usepackage{epsfig}

\def\Bra#1{\left<1>}
{\catcode`\|=\active\gdef\Braket#1{\left<\mathcode`\|"8000\let|\bravert {#1}\right>}}
\def\bravert{\egroup\,\vrule\,\bgroup}

\title{Tunable plasmons in atomically thin gold nanodisks}
\author{A.~Manjavacas}
\affiliation{IQFR - CSIC, Serrano 119, 28006 Madrid, Spain}
\altaffiliation{Current address: Department of Physics and Astronomy and Laboratory for Nanophotonics, Rice University, Houston, Texas 77005, United States}
\author{F.~J.~Garc\'{\i}a de Abajo}
\email{javier.garciadeabajo@icfo.es}
\affiliation{ICFO - Institut de Ciencies Fotoniques, Mediterranean Technology Park, 08860 Castelldefels (Barcelona), Spain}
\alsoaffiliation{ICREA - Instituci\'o Catalana de Recerca i Estudis Avan\c{c}ats, Pg. Llu\'{\i}s Companys 23, 08010 Barcelona, Spain}
\begin{document}

\begin{abstract}
The ability to modulate light at high speeds is of paramount importance for telecommunications, information processing, and medical imaging technologies. This has stimulated intense efforts to master optoelectronic switching at visible and near-infrared frequencies, although coping with current computer speeds in integrated architectures still remains a major challenge. As a partial success, midinfrared light modulation has been recently achieved through gating patterned graphene. Here we show that atomically thin noble metal nanoislands can extend optical modulation to the visible and near-infrared spectral range. We find plasmons in thin metal nanodisks to produce similar absorption cross-sections as spherical particles of the same diameter. Using realistic levels of electrical doping, plasmons are shifted by about half their width, thus leading to a factor-of-two change in light absorption. These results, which we substantiate on microscopic quantum theory of the optical response, hold great potential for the development of electrical visible and near-infrared light modulation in integrable, nanoscale devices.
\end{abstract}
\maketitle

\section{Introduction}

Surface plasmons, the collective oscillations of conduction electrons in metallic structures, allow us to confine light down to deep subwavelength volumes \cite{NH06}. Additionally, they couple strongly to electromagnetic fields \cite{HLC11}. Because of these properties, plasmons are excellent tools to engineer nanoscale devices for manipulating optical signals, without the limitation imposed by diffraction in far-field setups. This has triggered a number of applications in areas as diverse as ultrasensitive biosensing \cite{paper156}, improved photovoltaics \cite{AP10}, plasmon enhanced photodetection \cite{KSN11}, and photothermal cancer therapy \cite{NHH04}. The design of plasmonic structures with suitable spectral characteristics involves a careful choice of geometry and composition. In recent years, a vast amount of work has been devoted to producing nanostructures made of noble metals with controlled size and morphology, using in particular colloidal methods \cite{GPM08} and lithography \cite{SMC07}.

Despite these advances in the control over the {\it static} characteristics of plasmons, the dynamical modulation of their frequencies and spatial profiles remains ellusive, particularly in the visible and near-infrared (vis-NIR) parts of the spectrum. In this context, slow mild changes of the plasmon frequency have been produced by electrochemically injecting electrons in metal nanoparticles \cite{MPG06}, by electrically driving liquid crystals containing plasmonic particles \cite{CCC06_2}, and through controllable metamaterial designs \cite{BGK10,LZT12}. Magneto-optical modulation has also been explored to control plasmons in noble metal structures \cite{ACG13}. Hybrids of plasmonic and conductive oxides have been proposed \cite{FDA10,AAA11}, as well as colloids based on different materials \cite{CM14}. However, we still need to devise new methods to produce larger and faster control over plasmons, as required for nanoscale optical commutation and light modulation at high speeds.

Recently, the emergence of graphene \cite{CGP09} as a novel plasmonic material \cite{JBS09,paper176,GPN12} has opened up new paths towards the design of dynamically tunable plasmonic devices. Electrically doped graphene supports surface plasmons whose frequency can be efficiently varied by changing the level of doping \cite{paper196,FRA12,paper212}.
Consequently, the resulting modulation is intrinsically fast because it can be driven by charge-carrier injection using conventional electric gating technology. This promising material has been so far shown to support mid-infrared and lower-frequency plasmons \cite{JGH11,paper196,FRA12,paper212,BJS13}, while vis-NIR modes are being pursued by reducing the size of the structures \cite{paper214,paper215} and increasing the level of doping \cite{paper212}. The search for plasmon modulation in the vis-NIR is thus still ongoing, as these are spectral regions of utmost importance for sensing and optical signal processing technologies.

The origins of the excellent tunability of plasmons in graphene can be found in both the atomic thickness and the peculiar electronic structure of this material. The latter is characterized by a linear dispersion relation, which leads to a vanishing of the density of states at the Fermi level, so that a relatively small density of injected charge carriers produces substantial optical gaps in which collective plasmon modes emerge \cite{CGP09,paper176}. Although this unique feature cannot be easily transported to conventional plasmonic materials, such as gold, we can still mimic graphene plasmonics by going to atomically thin noble metals, whose optical response should be more susceptible to doping than traditional thicker layers. In particular, monolayer gold, the synthesis of which has been mastered for a long time in the context of surface science \cite{GB1981}, presents the advantage of having a plasma frequency compatible with the existence of plasmons in the vis-NIR \cite{JC1972}.

Here, we show that single-monolayer gold disks (SMGDs) with diameters of the order of $10\,$nm support surface plasmons with large cross-sections comparable to their geometrical areas. The frequencies of these excitations lie in the vis-NIR and can be efficiently modulated using attainable concentrations of doping charge carriers, which can be provided via electrical doping using for example backgating technology. We also analyze the optical response of periodic arrays of SMGDs, for which we predict an absorbance $\sim25\%$ for metal layer filling fractions $\sim40\%$.

\section{Results and Discussion}

\subsection{Electrically tunable optical response}

The system under study is depicted in Fig.\ 1a. We consider a gold nanodisk of diameter $D$, extracted from a single (111) atomic layer of gold. We take the thickness of the gold monolayer to be equal to the separation between (111) atomic planes in bulk gold ({\it i.e.}, $a_0/\sqrt{3}$, where $a_0=0.408\,$nm is the atomic lattice constant). Incidentally, our results are rather independent on the choice of disk thickness when this is small compared with the diameter, as long as the total valence charge is preserved (see Supplementary Fig.\ 6). As a first step in our analysis, we describe the optical response of a SMGD classically by modeling it as a thin disk described by a frequency-dependent homogeneous dielectric function $\epsilon(\omega)$. More precisely, we calculate the extinction cross-section $\sigma$ by solving Maxwell's equations using the boundary-element method (BEM) \cite{paper040}. Interestingly, for a diameter $D=20\,$nm, the cross section is dominated by a NIR plasmon at an energy $\sim1\,$eV and it exceeds the geometrical area of the disk (see left part of Fig.\ 1b).

It is convenient to separate the contribution from s-band electrons in the dielectric function as a Lorentzian term,
\begin{equation}
\epsilon\left(\omega\right)= \epsilon_{\rm b} - \frac{\omega^2_{\rm p}}{\omega\left(\omega+\ii\gamma\right)}, \label{Drude}
\end{equation}
where $\epsilon_{\rm b}$ accounts for the effect of {\it background} screening due to d-band electrons, $\hbar\omega_{\rm p}=9.06\,$eV is the classical plasmon energy associated with s valence electrons (see Methods), and $\hbar\gamma=71\,$meV is an inelastic width (we adopt this value of the damping throughout this work). As explained below, we introduce doping in the classical model by changing $\omega_{\rm p}$ in Eq.\ (1). In general, the description of the response of gold in the vis-NIR region including interband transitions requires to use either experimental data \cite{JC1972} or a sophisticated multi-Lorentzian model \cite{RDE98} for $\epsilon(\omega)$, which yields a $\omega$-dependent background $\epsilon_{\rm b}=\epsilon(\omega)+\omega^2_{\rm p}/[\omega(\omega+\ii\gamma)]$. However, we show in Fig.\ 1b that a simple Drude model for Eq.\ (1) (dashed curves), consisting in fixing $\epsilon_{\rm b}=9.5$ for all frequencies, produces a satisfactory level of accuracy at the observed relatively low disk-plasmon energies compared with the results obtained from tabulated optical data \cite{JC1972} (solid curves). Additionally, as the Drude model (i.e., constant $\epsilon_{\rm b}$) provides a natural connection with the quantum-mechanical approach described below, we use for disks it in what follows.

\begin{figure}
\begin{center}
\includegraphics[width=160mm,angle=0]{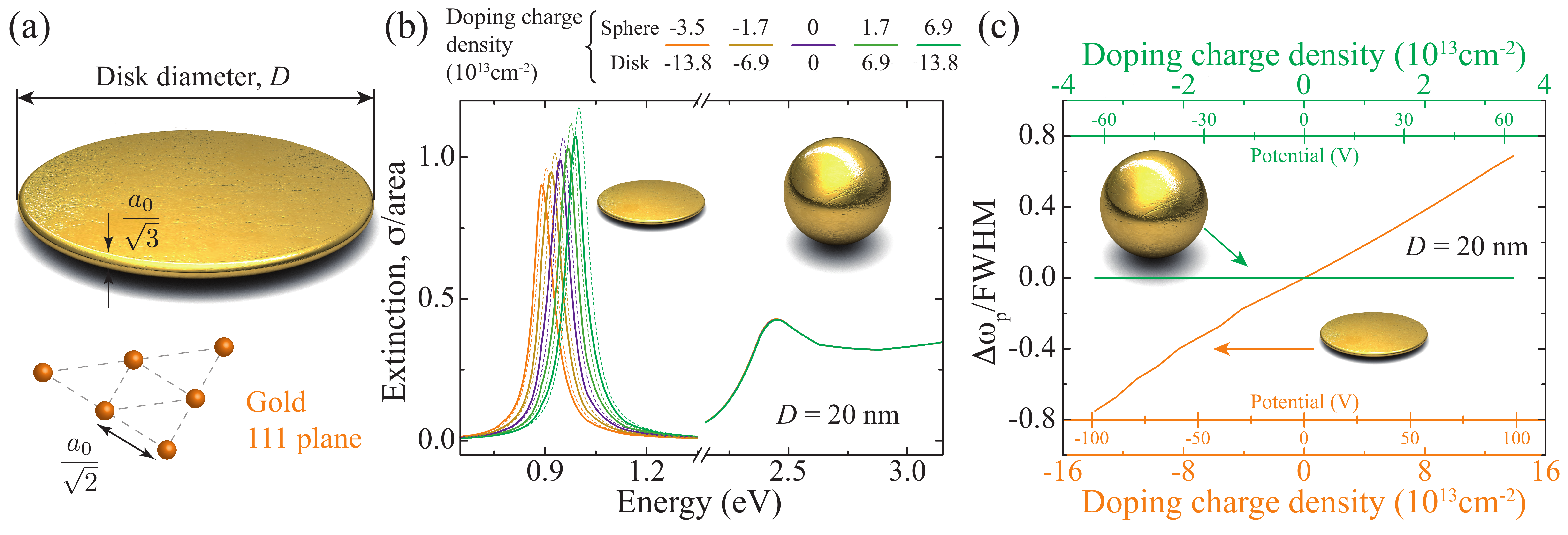}
\caption{{\bf Optical response and electrical tunability of single-monolayer gold disks.} (a) We consider a single-monolayer gold disk (SMGD) of diameter $D$ carved from a single (111) atomic layer. The disk thickness is $a_0/\sqrt{3}$, where $a_0=0.408\,$nm is the atomic lattice constant. (b) Extinction cross-section of a $D=20\,$nm SMGD for different doping charge-carrier densities (see upper legend). A doping density of $13.8\times10^{13}\,$cm$^{-2}$ corresponds to total of 440 additional charge carriers in the disk. For comparison, we also plot the cross section for a gold nanosphere of the same diameter and total doping charge, clearly showing an almost negligible degree of tunability. The particles are assumed to be homogeneously doped and described classically through the local dielectric function tabulated from measured optical data (solid curves). \cite{JC1972} Results obtained from a Drude dielectric function (Eq.\ (1)) are shown for comparison (broken curves). (c) Plasmon frequency shift relative to the width of the plasmon resonance for the disk (orange) and the sphere (green) of panel (b) as a function of doping density. The small scales indicate the potential at the disk/sphere surface for different disk doping densities.}\label{fig1}
\end{center}
\end{figure}

We consider next the effect of electrical doping. The spatial distribution of additional charge carriers depends on the doping configuration, as it can be for example homogenous for disks connected to a non-absorbing gate (e.g., ITO) or inhomogeneous in self-standing charged disks, although the plasmon energies and spatial profiles are expected to be similar in both cases based upon our experience with graphene disk plasmons \cite{paper194}. For simplicity, we assume homogeneously doped disks in what follows. The additional doping charge density $n$ adds up to the undoped s band density $n_0=m_{\rm e}\omega_{\rm p}^2/(4\pi e^2)\approx1.4\times10^{15}\,$cm$^{-2}$, which is rather close to the s-band areal electron density in neutral monolayer gold, $4/\sqrt{3}a_0^2$. The doping charge is thus introduced by changing the bulk plasma frequency to $\omega_{\rm p}=\left[\left(4\pi e^2/m_{\rm e}\right)(n_0+n)\right]^{1/2}$ in Eq.\ (1). Now, the addition of a moderate amount of doping electrons ($\sim5-10\%$ of $n_0$) results in significant blue shifts and increase in the strength of the plasmon resonance (\emph{cf.} purple and green curves of Fig.\ 1b). Obviously, the injection of similar amounts of holes produces the opposite effects (Fig.\ 1b, orange curve). The small thickness of the disk is a key factor in producing such dramatic modifications in the optical response using realistic doping densities. In fact, repeating this operation with a gold nanosphere of the same diameter, we also observe a prominent plasmon (Fig.\ 1b, $\sim2.5\,$eV region), but it remains unchanged when adding similar amounts of doping charges. In the sphere, the doping charges pileup in the outermost atomic layer \cite{LK1970_2}, but this produces the same extinction cross-section as if the charges where homogenously distributed over its entire volume, and therefore, the change in bulk charge density is substantially reduced with respect to the disk.

Figure\ 1c compares the modulation of the nanodisk and the sphere. In particular, we plot the frequency shift normalized to the full width at half maximum (FWHM) for the plasmon resonance as a function of doping charge density. In contrast to the negligible tunability of the sphere, the disk allows shifts comparable to the FWHM to be electrically induced. Incidentally, the doping densities here considered produce realistic values of the electrostatic potential at the surface of these nanoparticles (Fig.\ 1c, small scales), indicating that they are compatible with currently available backgating technology \cite{paper212}. 

\begin{figure}
\begin{center}
\includegraphics[width=160mm,angle=0]{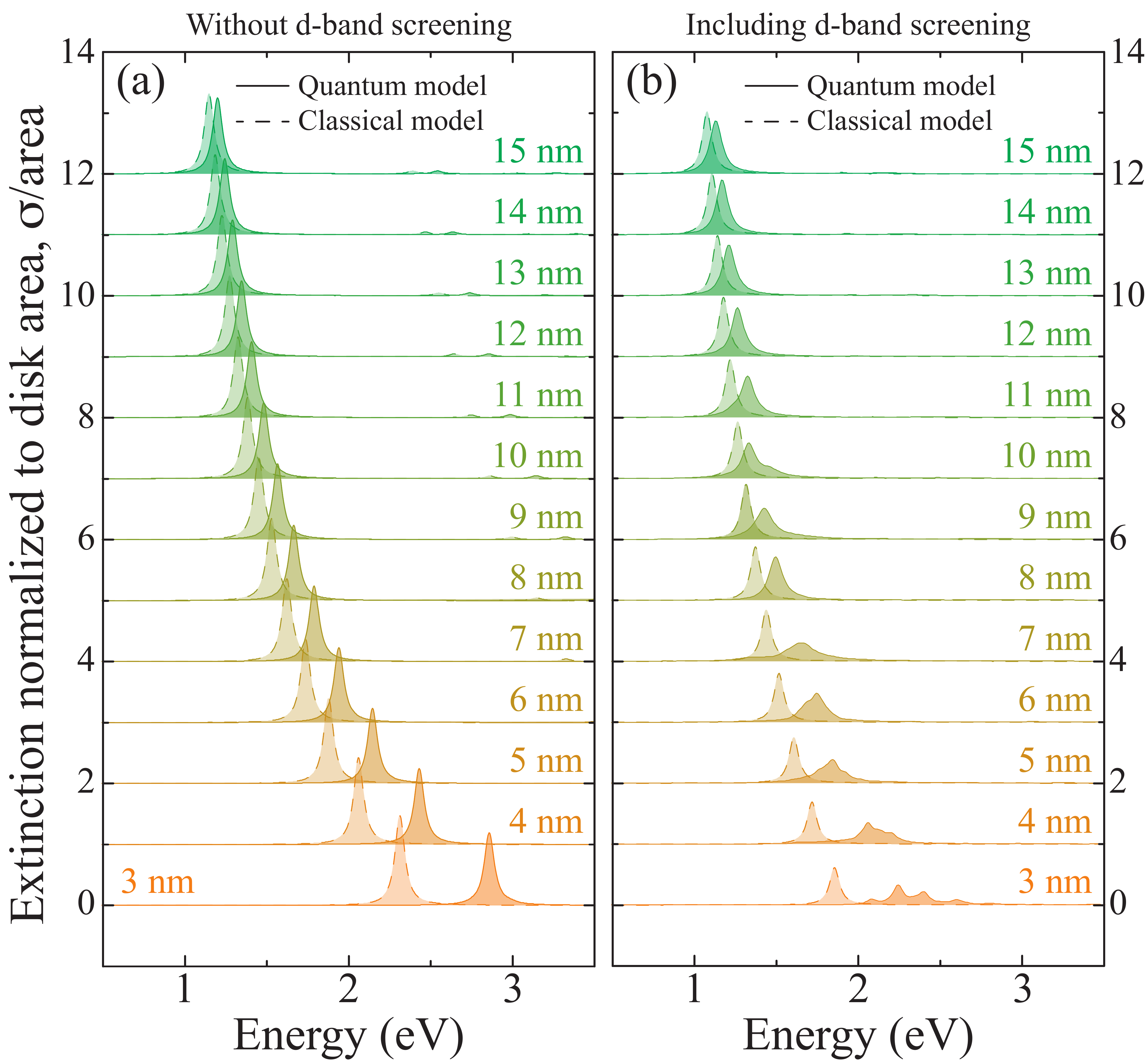}
\caption{{\bf Optical response of individual single-monolayer gold disks.} We plot the extinction cross-section normalized to the geometrical area for different diameters $D$, calculated using the quantum model (solid curves) and a classical description (dashed curves). Results obtained with and without inclusion of d-band screening are shown in (b) and (a), respectively.}\label{fig2}
\end{center}
\end{figure}

\subsection{Quantum-mechanical effects}

For nanoparticles of only a few nanometers in size, the above classical description fails to account for spatial dispersion and quantum confinement effects \cite{ZPN09,SHE12}, which generally require models based on a quantum-mechanical treatment of valence electrons and their interactions. Here, we use the random-phase approximation (RPA) to calculate the optical response of SMGDs (see Methods) within the electrostatic approximation, which should be rather accurate given the small sizes of the particles under consideration. This allows us to determine the validity of the classical approach and explore the response of nanodisks with smaller diameters.

We use particle-in-a-box states to describe independent s-band electrons. The cylindrical box has the same dimensions as in the classical calculations (see above) and it is surrounded by an infinite potential. The RPA susceptibility is then evaluated using these electron states to obtain the induced charge density, which in turn allows us to compute the optical extinction of the disk. Additionally, we model screening due to d-band electrons through an array of point dipoles placed at the atomic positions in the (111) layer and with polarizability adjusted to render an effective background permittivity $\epsilon_{\rm b}$ in the bulk material (see Methods). Apart from the relative position of these dipoles with respect to the disk center, our quantum description only depends on the three same parameters as the classical Drude theory (i.e., $\epsilon_{\rm b}$, $\omega_{\rm p}$, and $\gamma$).

A major assumption we are making is that $\gamma$ takes the same value as in the bulk metal. We use this as a reasonable estimate because s-band electrons are rather delocalized along directions parallel to the layer (i.e., similar to the bulk), while they are narrowly confined to the ground state across the transversal direction, so that plasmons result from in-plane motion. However, the actual value of $\gamma$ might depend on the detailed coupling of valence electrons to impurities and to the atomic lattice. Concerning d-band screening, our effective dipoles approach should provide a more realistic description than a homogeneous polarizable background. Although the discreteness of the dipole lattice can have strong effects in small islands, we find converged results for large islands, which are independent of the alignment of the dipole lattice relative to the disk center.

Figure\ 2 shows the extinction cross-section normalized to the disk area for SMGDs of different diameters ranging from $3$ to $15\,$nm. The main conclusions from this figure are as follows: (1) the extinction cross-sections are of the order of the disk area; (2) the plasmon energy increases with decreasing diameter $D$, exhibiting an approximate $\propto\sqrt{D}$ dependence, similar to what one finds in graphene nanodisks \cite{paper212}; (3) quantum calculations produce energies above those predicted by classical theory, as well as broader plasmon peaks, but the discrepancy between the two models decreases with increasing diameter; (4) in the absense of d-band screening (Fig.\ 2a, obtained by setting $\epsilon_{\rm b}=1$), both levels of description give rise to smooth plasmon peaks, in contrast to the quantum results obtained when d-band screening is switched on (Fig.\ 2b, with $\epsilon_{\rm b}=9.5$); (5) d-band screening also leads to a redshift of the plasmons, which is more pronounced at small sizes. Incidentally, the induced charge associated with the plasmon exhibits a dipolar profile dressed with radial oscillations mimicking those of Friedel oscillations, which are particularly intense for small diameters (see Supplementary Fig.\ 7).

Similar blue shifts with respect to classical local theory are also found in small noble metal particles \cite{KV95}, the origin of which is a combination of nonlocal and quantum effects, particularly due to the surface spill out of s electrons beyond the polarizable background of d-band electrons. In simple metals such as aluminum, the spill out produces smaller electron densities near the surface, and consequently, also smaller surface plasmon frequencies. In contrast, in noble metals, the spill out results in a weaker interaction with the localized d electrons, and thus, it leads to an increase in the observed frequency, which overcomes the redshift due to the smaller electron density. \cite{L93} We incorporate here the finite extension of s electrons across the normal direction, combined with the localization of the effective d-band dipoles, leading to similar blue shifts. Interestingly, our quantum model predicts splitting of the plasmon into multiple peaks for small disks when d-band screening is included (see for example the $D=3\,$nm spectrum in Fig.\ 2b). The presence of these peaks, which are rapidly coalescing into a single plasmon resonance for $D>8\,$nm, is a manifestation of the discrete character of the interaction with d-band electrons, which is more pronounced for small $D$'s. The jumps observed in the FWHM also shares a similar origin. It should be noted that these effects could be sensitive to the exact form of the s-electron transversal wave function in the smallest islands under consideration, which require a more atomistic analysis, based for example upon density-functional theory \cite{ORR02,LM14}. Likewise, the spectra for the smallest disks depend on the alignment of the d-band dipole lattice relative to the edges. In practice, 1D faceting of the edges becomes an important source of anisotropy, which can contribute to broaden the spectra for $D<5\,$nm.

\begin{figure}
\begin{center}
\includegraphics[width=160mm,angle=0]{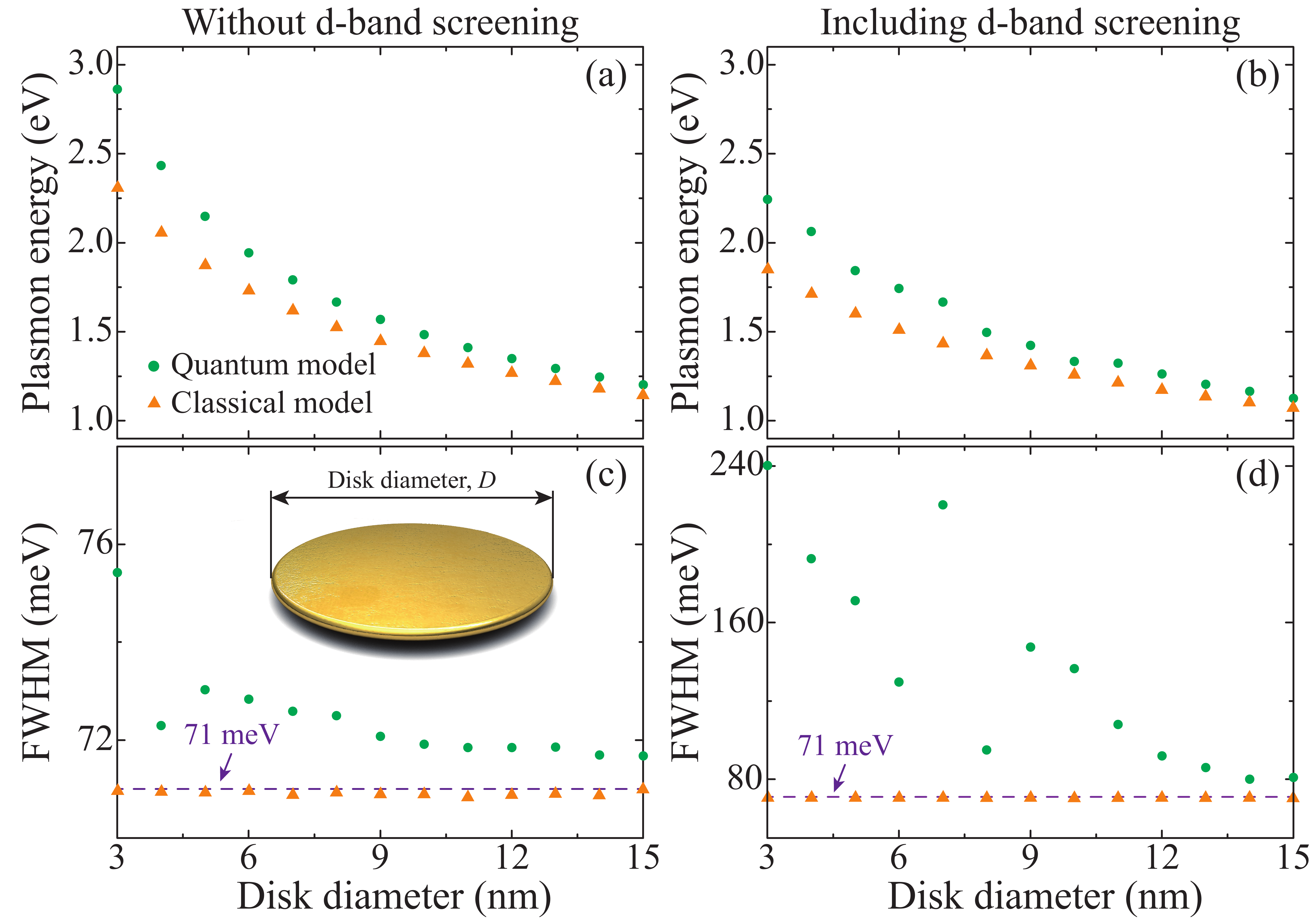}
\caption{{\bf Comparison of quantum and classical plasmon energies and widths.} Energy (a,b) and FWHM (c,d) of the plasmonic resonance of individual SMGDs as a function of disk diameter, calculated from quantum (green circles) and classical (orange triangles) models. Results obtained with and without inclusion of d-band screening are shown in (b,d) and (a,c), respectively. The dashed curves in (c,d) indicate the intrinsic broadening $\hbar\gamma=71\,$meV introduced in the Drude formula (Eq.\ (1)) and in the RPA susceptibility (Eq.\ (4)).}\label{fig3}
\end{center}
\end{figure}

The convergence of the quantum model to the classical description for increasing diameter is clearly observed in Fig.\ 3, which summarizes the plasmon energies and widths observed in the spectra of Fig.\ 2. Here, we define the FWHM as the frequency interval around the peak maximum that contains half of its area; this definition coincides with the standard FWHM for individual Lorentzian resonances, but it can be applied to multiple resonances as well to yield an overall width (in particular to the lower quantum-model spectra of Fig.\ 2b). Within the electrostatic limit under consideration, the FWHM predicted by the classical model is independent of diameter and equals the damping energy $\hbar\gamma=71\,$meV (see Eq.\ (1)). In contrast, the quantum model leads to a significant increase in the FWHM for small diameters, essentially as a consequence of Landau damping, which involves inelastic decay of plasmons to electron-hole pairs for momentum transfers $\sim\omega/v_{\rm F}$, where $v_{\rm F}$ is the Fermi velocity ($\sim10^6$m\,s$^{-1}$, see Supplementary Fig.\ 8a). As the momentum transfer provided by the breaking of translational invariance in a disk is $\sim2\pi/D$, the onset of Landau damping is expected to occur at $D\sim2\pi v_{\rm F}/\omega\sim4\,$nm, in qualitative agreement with the results shown in Fig.\ 3c,d. An intuitive estimate can be stablished from the electron mean free path $v_{\rm F}/\gamma\sim10\,$nm, which determines the rate of collisions with the edges ({i.e.}, events that provide the noted momentum), and is also in agreement with the trends observed in Fig.\ 3c,d, although the value of $\gamma$ regarded as a parameter simply produces an additional contribution to the FWHM that is independent of $D$, and the ultimate origin of broadening for small sizes can be found in Landau damping.

\begin{figure}
\begin{center}
\includegraphics[width=160mm,angle=0]{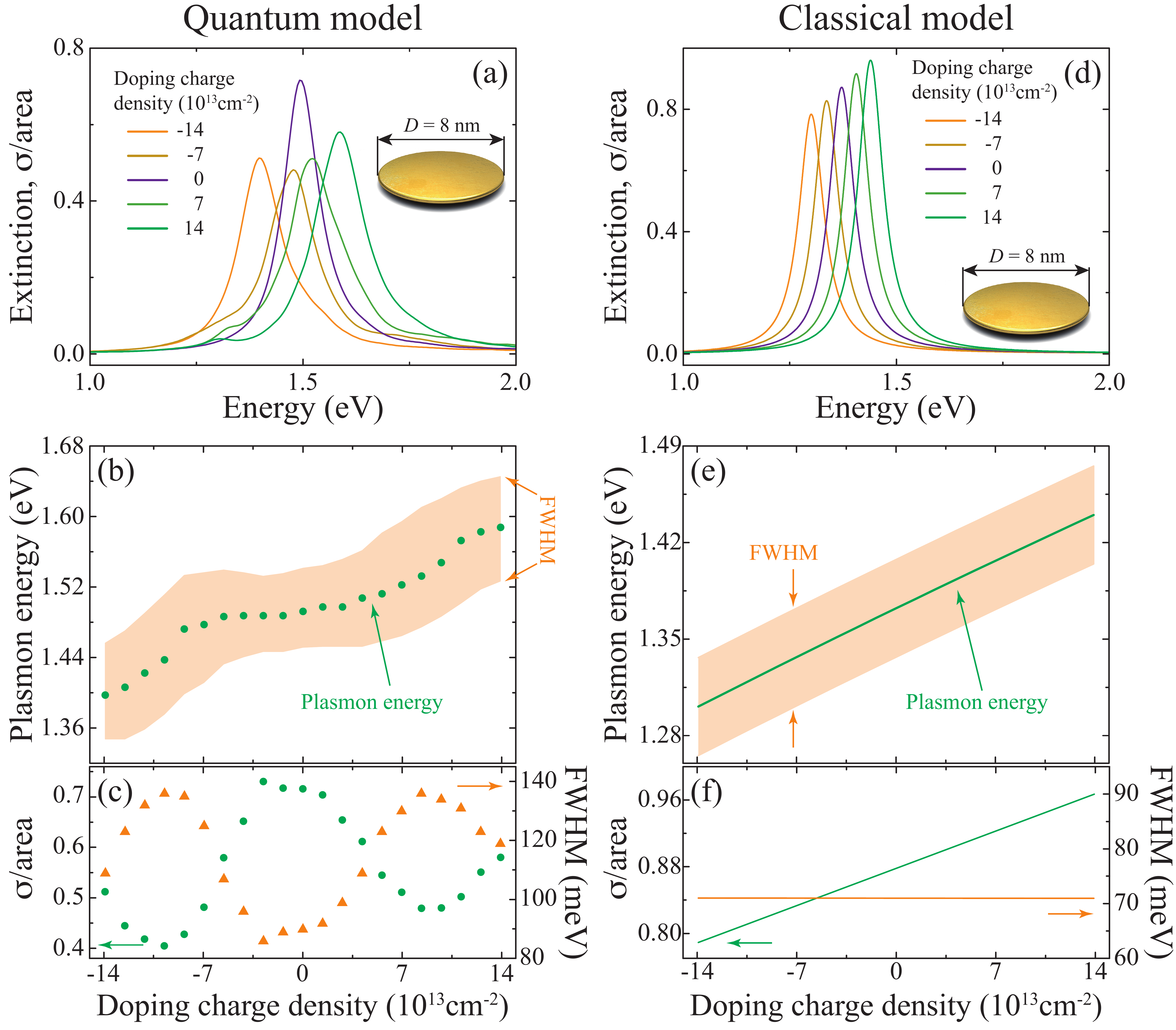}
\caption{{\bf Quantum vs classical analysis of the electrically tunable optical response.} (a,d) Extinction cross-section normalized to the geometrical area for a $D=8\,$nm single-monolayer gold disk calculated with different doping charge densities. (b,e) Plasmon energy as a function of doping charge density. We indicate the FWHM of the plasmon resonances as shadowed regions. (c,f) Optical extinction cross-section at the plasmon peak energy (green curves and symbols, left scale) and FWHM (orange, right scale) as a function of doping density. Quantum mechanical calculations (a-c) are compared with classical results (d-f), including d-band screening in all cases.}\label{fig4}
\end{center}
\end{figure}

As discussed in Fig.\ 1, the optical response of SMGDs can be modified through the addition of small amounts of charge carriers to relatively large disks ($D=20\,$nm), for which classical theory is rather accurate (see Figs.\ 2 and 3). Using smaller SMGDs, we obtain qualitatively similar results as for larger disks (see Fig.\ 4 for an analysis of a $D=8\,$nm doped disk). Given the small disk size, we compare classical (Fig.\ 4d-f) and quantum (Fig.\ 4a-c) results, showing again a blue shift and plasmon broadening in the latter relative to the former.

In contrast to the nearly linear plasmon shift with doping charge density predicted by classical theory (Fig.\ 4e), the quantum model leads to initially smaller modulation at low doping (Fig.\ 4b), which increases to a faster pace than the classical results for higher doping. This nonlinear dependence of the plasmon energy on the doping density could be exploited for improved light modulation by operating around a highly doped SMGD configuration. In particular, the plasmon shift can be as large as the FWHM when the density of s-band electrons is changed by $\pm(5-10)\%$.

Interestingly, the nonlinear plasmon shifts observed in the quantum model becomes oscillatory when examining the maximum extinction cross-section and the FWHM (Fig.\ 4c). The oscillations of these two quantities are out of phase, as required to satisfy the $f$-sum rule \cite{PN1966}, and can be traced back to the discreteness of the electronic energies. Importantly, in all cases the maximum cross-section is of the order of the disk area (Fig.\ 4c,f), thus providing good coupling to light for potential applications to modulation devices.

\begin{figure}
\begin{center}
\includegraphics[width=160mm,angle=0]{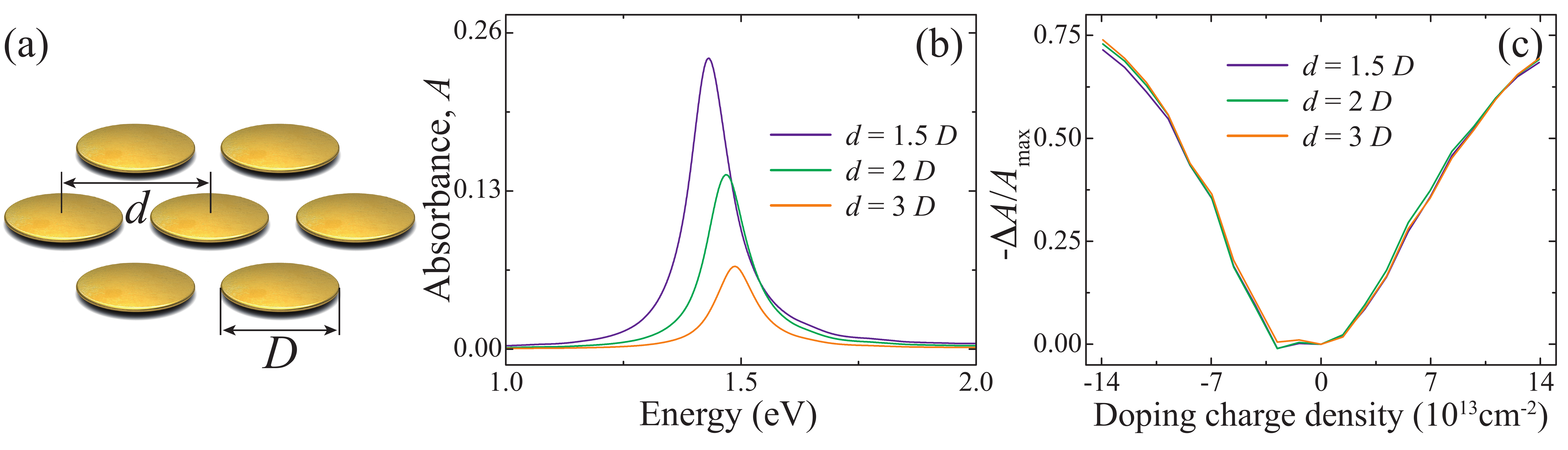}
\caption{{\bf Electrical modulation of the absorbance of an hexagonal periodic arrangement of single-monolayer gold disks.} (a) Scheme of the system under study. (b) Absorbance spectrum of undoped nanodisks (diameter $D=8\,$nm) for different values of the array spacing $d$. (c) Modulation of the absorbance relative to the undoped state as a function of doping charge density for different array spacings.}\label{fig5}
\end{center}
\end{figure}

\subsection{Periodic arrangement of single-monolayer gold nanodisks}

The large optical strength and degree of electrical tunability discussed above for SMGDs can be exploited to modulate light that is either transmitted or reflected by a periodic array of such structures. We consider an hexagonal array of $D=8\,$nm disks in Fig.\ 5 with different values of the array spacing $d$. Given the large mismatch between $D$ and the resonant light wavelength ($\sim830\,$nm), we approximate the disks as point dipoles of polarizability extracted as explained in Methods (see also Supplementary Fig.\ 9, where we show that higher-order multipoles play only a small role for the relative distances under consideration). Following previous analytical methods \cite{paper182} to compute the absorbance $A$, we find remarkably large values (e.g., $A=25\%$ for $d=1.5\,D$, see Fig.\ 5b), given the small amount of gold in the structure (sub-monoatomic layer film). The fractional change in absorbance driven by electrical doping (Fig.\ 5c) is rather independent of lattice spacing and reaches $\sim70\%$ for a $10\%$ variation in the s-band electron density. The potential of patterned gold monolayers for electro-optical modulation in the NIR is thus excellent.

\section{Conclusions and Perspectives}

In summary, we have simulated both classically and quantum-mechanically the plasmonic response and performance in electro-optical modulation of gold nanodisks carved from a single (111) atomic layer. Our RPA calculations incorporate the wave functions of free s valence electrons evolving in a circular box, as well as an adjusted distribution of dipoles to account for d-band screening. Despite the atomic thickness of the disks, this quantum-mechanical description converges smoothly to the results of classical dielectric theory, based upon the bulk, frequency-dependent dielectric function of gold. This is a remarkable result, which can be intuitively understood from the fact that the electron current associated with the plasmons flows along the gold layer, and thus, it is rather insensitive to electron confinement within the small film thickness. Nontheless, nonlocality plays a crucial role, leading to strong plasmon blue shifts, as well as splitting due to the complex interaction with the d band. We estimate that nonlocal effects become dominant when the disk diameter is below $\sim10\,$nm.

Remarkably, the disks interact strongly with light, giving rise to extinction cross-sections exceeding their geometrical areas in the vis-NIR. We have also shown that the optical response of SMGDs can be efficiently modulated through the addition or removal of realistic concentrations of doping charge carriers using for example gating technology. In particularly, periodic patterns of monolayer gold appear to be a suitable solution for combining strong plasmonic response and high doping, for example using an electrical backgate, because the average charge density of the layer is simply determined by capacitor theory for a fixed distance from the gate, and thus, the actual doping charge density in the metal scales with the inverse of the areal filling fraction occupied by the gold. Similar results are expected for films consisting of only a few atomic layers, although the degree of modulation is then reduced because the doping charge has to be shared across the increased thickness. Other plasmonic metals such as silver and copper should find similar degree of tunability (see Supplementary Fig.\ 10). In particular, the small plasmon width of silver compared with gold makes it an attractive candidate to drive plasmon shifts well beyond the FWHM. Additionally, the lower d-band screening in this material should result in higher plasmon energies, reaching the visible in small disks, or equivalently, the NIR for larger disk diameters.

It should be stressed that, while the synthesis of single-layer gold is a mature field \cite{GB1981}, the fabrication of laterally confined thin gold nanostructures represents a technical challenge, which could benefit from advances in lithography and self-assembly. Alternatively, one could use a continuous gold layer, which also exhibits large electrical tunability of its propagating plasmons (see Supplementary Fig.\ 11), coupled to external light by decorating it with dielectric colloids in order to bridge the light-plasmon momentum mismatch (i.e., this is essentially what nanostructuration does in the SMGDs that we study above). The resulting planar structures hold great potential for light modulation at vis-NIR frequencies, which could be the basis of a new generation of electrically tunable optical devices with applications ranging from sensing to nanoscale spectroscopy.

\section{Methods}

\subsection{Quantum-mechanical RPA simulations}

We consider small disk sizes compared with the light wavelength, so that we work in the electrostatic limit. Within this approximation, assuming an overall monochromatic time dependence $\ee^{-\ii\omega t}$ with frequency $\omega$, the induced charge density $\rho^{\rm ind}$ can be expressed in terms of the self-consistent potential $\phi$ as
\begin{align}
\rho^{\rm ind}(\rb,\omega)=\int d^3\rb'\chi^0(\rb,\rb',\omega)\phi(\rb',\omega)\equiv\chi_0\cdot\phi,\label{rho}
\end{align}
where $\chi^0$ is the noninteracting susceptibility associated with the s valence electrons of gold, and the last identity defines a matrix notation in which matrix multiplication involves integration over space coordinates. We obtain $\chi^0$ within the RPA\cite{PN1966}, in which a one-electron picture is assumed and only individual electron-hole pair excitations are explicitly considered. We further approximate the wave functions of valence electrons by the solutions of a cylindrical box with the same diameter as the nanodisk and a thickness corresponding to the separation between (111) atomic planes in bulk gold
(\emph{i.e.}, $a_0/\sqrt{3}\approx0.236\,$nm, see Fig.\ 1a). More precisely,
\begin{align}
\psi_{lm}\left(\textbf{r}\right)=N_{lms}J_{m}\;\left(Q_{lm}R\right)\ee^{\ii m\varphi}g_1\left(z\right),\label{psi}
\end{align}
where $N_{lms}$ is a normalization constant, $Q_{lm}=2\zeta_{lm}/D$, $\zeta_{lm}$ is the $l^{\rm th}$ zero of the Bessel function $J_m$, and $g_1\left(z\right)=\sin\left(\pi\sqrt{3}z/a_0\right)$ yields the dependence on the coordinate $z$ normal to the disk. For simplicity, we are assuming that the $z$ dependence is separable in the wave function, so that electron diffraction effects at the disk edges are not important. Moreover, we assume that the electrons remain in the ground state of the vertical cavity of thickness $a_0/\sqrt{3}$, which is a reasonable approximation if we consider that the first excited state ({\it i.e.}, $g_2(z)=\sin\left(2\pi\sqrt{3}z/a_0\right)$) lies $\sim20\,$eV above $g_1$, well beyond the Fermi and vacuum levels.

With the wave functions of Eq.\ (3), we can write the susceptibility as
\begin{align}
\chi^0\left(\textbf{r},\textbf{r}',\omega\right)=\frac{2e^2}{\hbar}\sum_{l,l',m,m'}\left(f_{l'm'}-f_{lm}\right)
\frac{\psi_{lm}\left(\textbf{r}\right)\psi^{\ast}_{lm}\left(\textbf{r}'\right)\psi^{\ast}_{l'm'}\left(\textbf{r}\right)\psi_{l'm'}\left(\textbf{r}'\right)}
{\omega-\varepsilon_{lm}+\varepsilon_{l'm'}+\ii\gamma/2},\label{chi0}
\end{align}
where spin degeneracy is simply included through an overall factor of 2, $\hbar\varepsilon_{lm}=\hbar^2 Q^2_{lm}/2 m_{\rm e}$ is the energy of state $\psi_{lm}$ (notice that the energy associated with $z$ motion cancels out in Eq.\ (4), so we disregard it), $\gamma$ is an intrinsic relaxation time, which we take from a Drude fit (Eq.\ (1)) to measured optical data \cite{JC1972} ($\hbar\gamma=71\,$meV), and $f_{lm}=\left\{\exp\left[\left(\varepsilon_{lm}-E_{\rm F}\right)/k_{\rm B}T\right]+1\right\}^{-1}$ is the Fermi-Dirac distribution function, evaluated here at $T=0$. The method using to fill the energy levels in the disk is discussed in the Supplementary Fig.\ 8.

The total potential $\phi$ is the sum of the external potential $\phi^{\rm ext}$ and the potential produced by the induced charges
\begin{align}
\phi=\phi^{\rm ext}+v\cdot\rho^{\rm ind},\label{phi}
\end{align}
where $v(\rb-\rb')=1/|\rb-\rb'|$ is the Coulomb interaction. From here and Eq.\ (2), we solve the induced charge density as
\begin{align}
\rho^{\rm ind}=\chi^0\cdot\left(1-v\cdot\chi^0\right)^{-1}\cdot\phi^{\rm ext}.\label{rho2}
\end{align}
As the polarization along the direction normal to the disk is expected to be negligible, we focus on parallel components and write $\phi^{\rm ext}=-R\ee^{\ii\varphi}$ (i.e., we focus on solutions with $m=1$ azimuthal symmetry), where $\varphi$ is the azimuthal angle of $(x,y)$ and $R=\sqrt{x^2+y^2}$. This allows us to obtain the in-plane polarizability by calculating \[\alpha\left(\omega\right)=\frac{1}{2}\int d^3\textbf{r}\;R\ee^{-\ii\varphi}\;\rho^{\rm ind}\left(\textbf{r},\omega\right).\] Finally, the extinction cross-section is obtained from \[\sigma(\omega)=\left(4\pi\omega/c\right)\mbox{Im}\left\{\alpha\left(\omega\right)\right\}.\] 

\subsection{Inclusion of d-band screening}

Deeper electrons in the d band are relatively localized in the gold atoms, and therefore, we model them by assuming a background of polarizable point particles at the atomic positions in the (111) layer (see lower part of Fig.\ 1a). The polarizability $\alpha_{\rm b}$ of these particles is adjusted to fit the experimentally measured bulk dielectric function of gold $\epsilon_{\rm exp}$. That is, if we subtract the Drude s-band contribution from $\epsilon_{\rm exp}$ (see Eq.\ (1)), we obtain the background permittivity $\epsilon_{\rm b}=\epsilon_{\rm exp}+\omega_{\rm p}^2/\omega(\omega+\ii\gamma)$, where $\omega_{\rm p}^2=4\pi e^2 n_0/m_{\rm e}$ is determined by the s-band electron density $n_0=4/a_0^3\approx5.9\times10^{28}\,$m$^{-3}$. This yields $\hbar\omega_{\rm p}\approx9.01\,$eV, which is slightly different from the best fit of Eq.\ (1) to measured data \cite{JC1972} (9.06\,eV), from which we also find $\epsilon_{\rm b}=9.5$. Now the Clausius-Mossotti relation\cite{AM1976} leads to
\begin{align}
\alpha_{\rm b}=\frac{3}{4\pi n_0}\;\frac{\epsilon_{\rm b}-1}{\epsilon_{\rm b}+2}.\nonumber
\end{align}
Using dyadic notation, the susceptibility tensor of the background dipoles reduces to $\chi_{\rm b}^0(\rb,\rb')=\sum_j\overleftarrow{\nabla}\cdot\alpha_{\rm b}\delta(\rb-\rb_j)\delta(\rb'-\rb_j)\cdot\overrightarrow{\nabla}'$, where $j$ runs over the positions of the metal atoms, whereas $\overleftarrow{\nabla}$ ($\overrightarrow{\nabla}'$) acts on $\rb$-dependent ($\rb'$-dependent) functions to the left (right) of operator $\chi_{\rm b}^0$. As the charge induced through both s and d bands contribute together to the full potential, we can rewrite Eq.\ (2) as
\begin{align}
\rho^{\rm ind}=\left(\chi^0+\chi^0_{\rm b}\right)\cdot\phi\nonumber
\end{align}
to take into account the effect of d-band screening.
Using this expression together with Eq.\ (5), the total induced charge density becomes
\begin{align}
\rho^{\rm ind}=\left(\chi^0+\chi^0_{\rm b}\right)\cdot\left[1-v\cdot\left(\chi^0+\chi^0_{\rm b}\right)\right]^{-1}\cdot\phi^{\rm ext},\label{rho3}
\end{align}
from which we calculate the disk polarizability and the extinction cross-section as explained above.

\pagebreak

\section{Supplementary Figures}

\begin{figure}
\begin{center}
\includegraphics[width=130mm,angle=0]{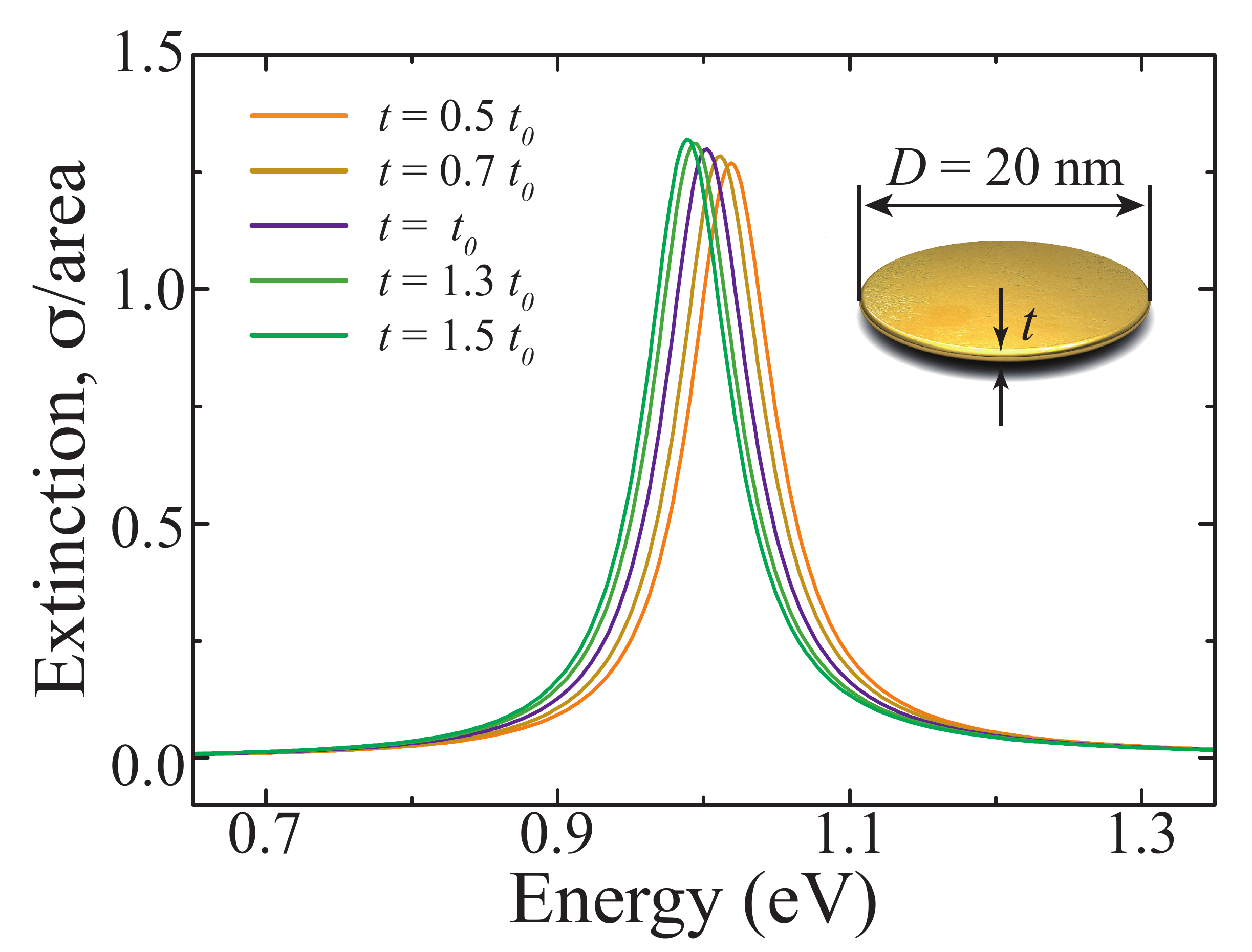}
\caption{{\bf Plasmon dependence on disk thickness.} We show classical calculations for disks of different thicknesses ($t$ in units of $t_0=a_0/\sqrt{3}=0.236\,$nm, see legend) and $D=20\,$nm in diameter. The valence electron density is adjusted to have the same number of electrons in all cases. These results show that the plasmon energies and absorption profiles are rather independent on disk thickness, provided this is small compared with the diameter, which corroborates the robustness of our calculations with respect to the choice of film thickness.}\label{figS1}
\end{center}
\end{figure}

\begin{figure}
\begin{center}
\includegraphics[width=140mm,angle=0]{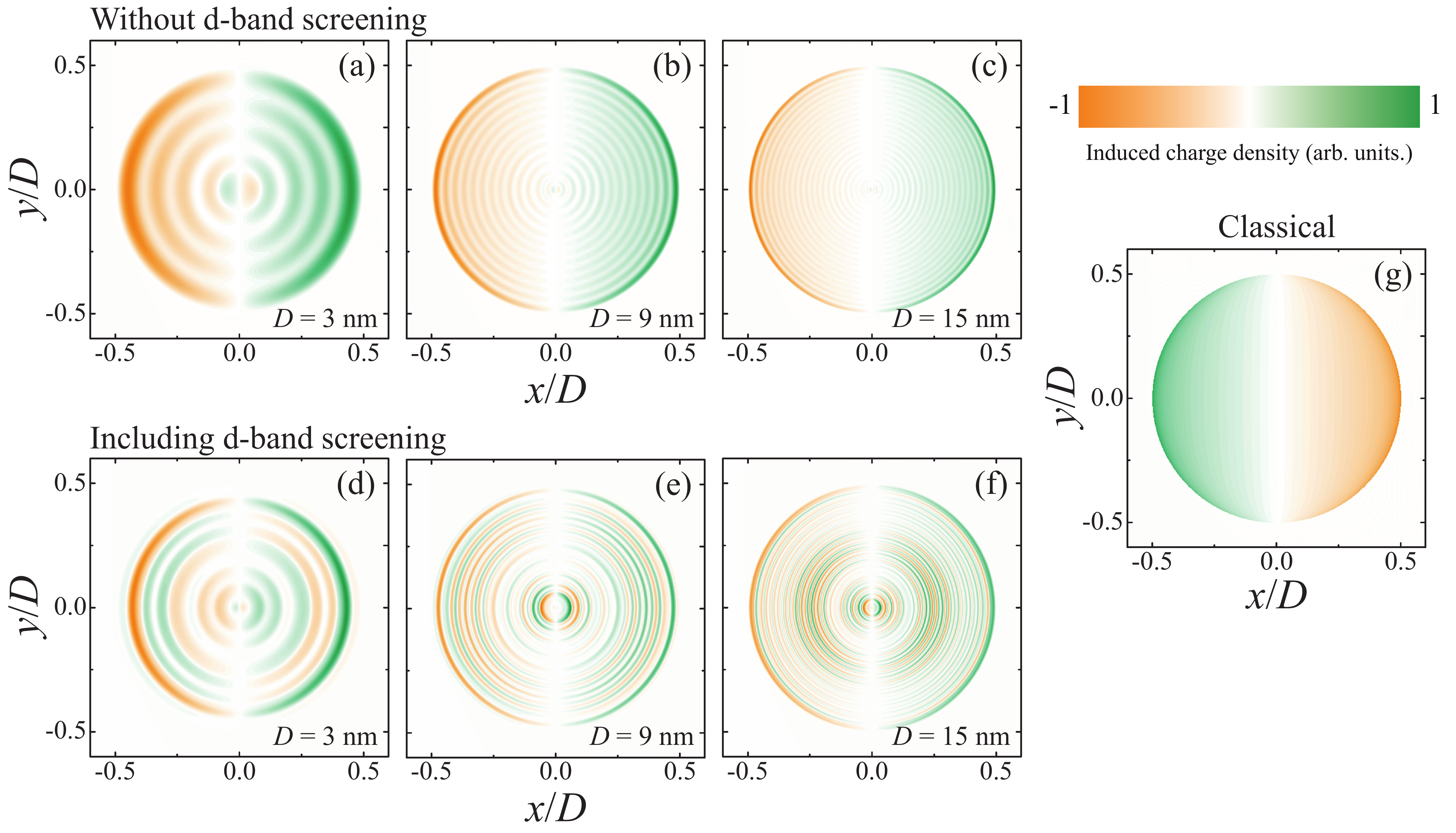}
\caption{{\bf Maps of induced charge density corresponding to the lowest-order dipolar plasmon of single-monolayer gold disks.} We consider disks of different diameters $D$ and compare the density obtained from our quantum mechanical description (a-c,d-f) with classical, local theory (g). The upper (lower) row shows results calculated without (with) inclusion of d-band electron screening. The classical calculation (g) is obtained from the solution of Poisson's equation for a disk described by a resonant permittivity \cite{paper212} $\epsilon\approx1-D/t$, where $t$ is the disk thickness. The dipolar patttern exhibits radial oscillations of a period similar to the Friedel oscillations. The inclusion of d-band screening (Fig.\ 7d-f) results in more complex patterns driven by the discrete character of the background dipoles. In all cases considered, the dipolar character is clearly preserved and the induced charge accumulates at the border of the nanodisk as the diameter increases, thus approaching the behavior of a classical, local description of the disk.}\label{figS2}
\end{center}
\end{figure}

\begin{figure}
\begin{center}
\includegraphics[width=110mm,angle=0]{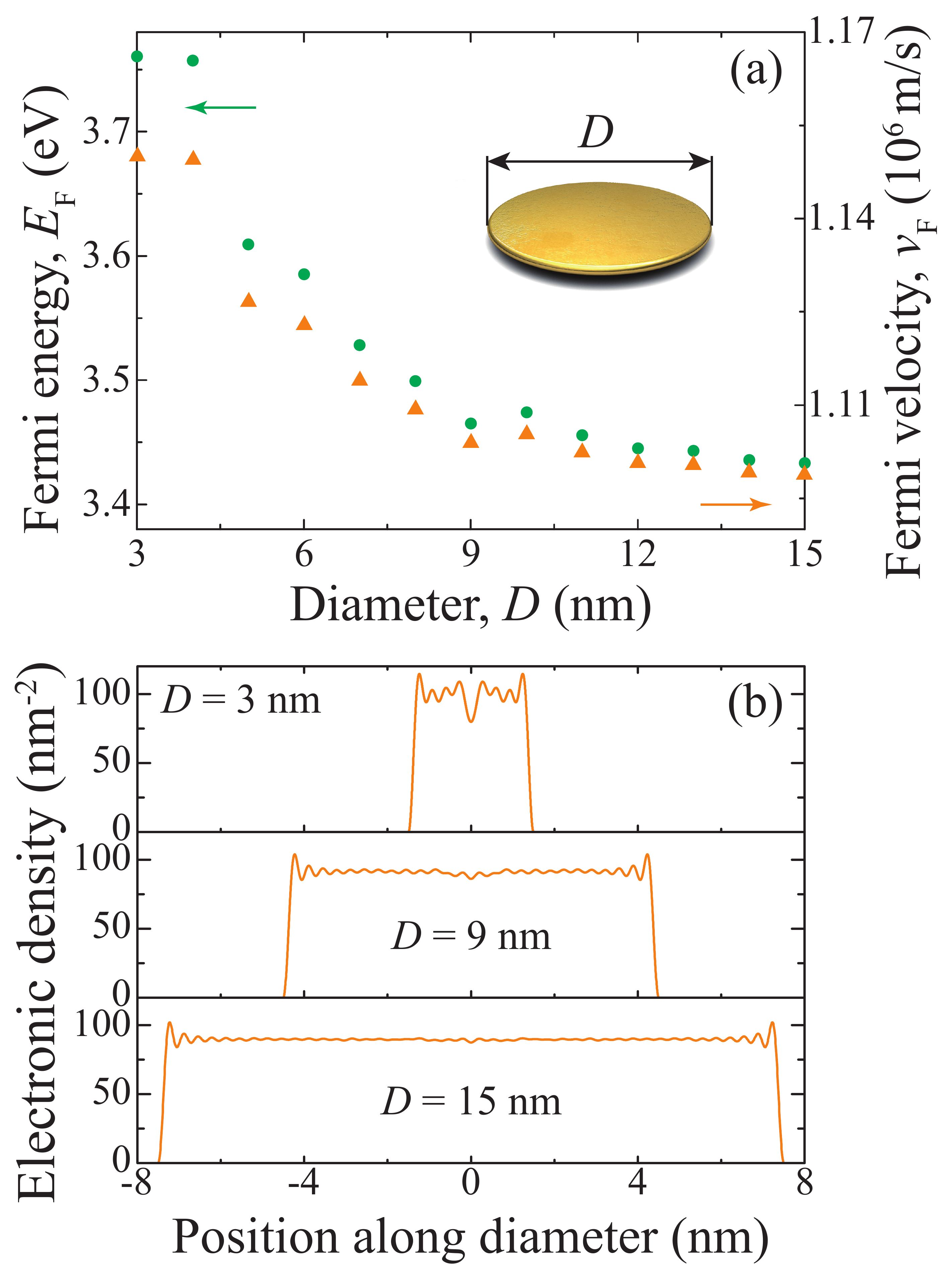}
\caption{{\bf Filling of electron energy levels in a metallic disk.} (a) Fermi energy $E_{\rm F}$ relative to the bottom of the parabolic s band (left scale) and Fermi velocity $v_{\rm F}=\sqrt{2E_{\rm F}/m_{\rm e}}$ (right scale) as a function of single-monolayer gold disk (SMGD) diameter. We consider a single (111) atomic layer of an fcc metal of lattice constant $a_0$ (e.g., $a_0=0.408\,$nm in gold). In practice, a single-layer disk of diameter $D$ has $\sim(\pi/\sqrt{3})(D/a_0)^2$ electrons that fill states of increasing energy up to a level that defines the Fermi energy $E_{\rm F}$.
(b) Unperturbed valence electron density profiles for SMGDs of different diameters.}\label{figS3}
\end{center}
\end{figure}

\begin{figure}
\begin{center}
\includegraphics[width=120mm,angle=0]{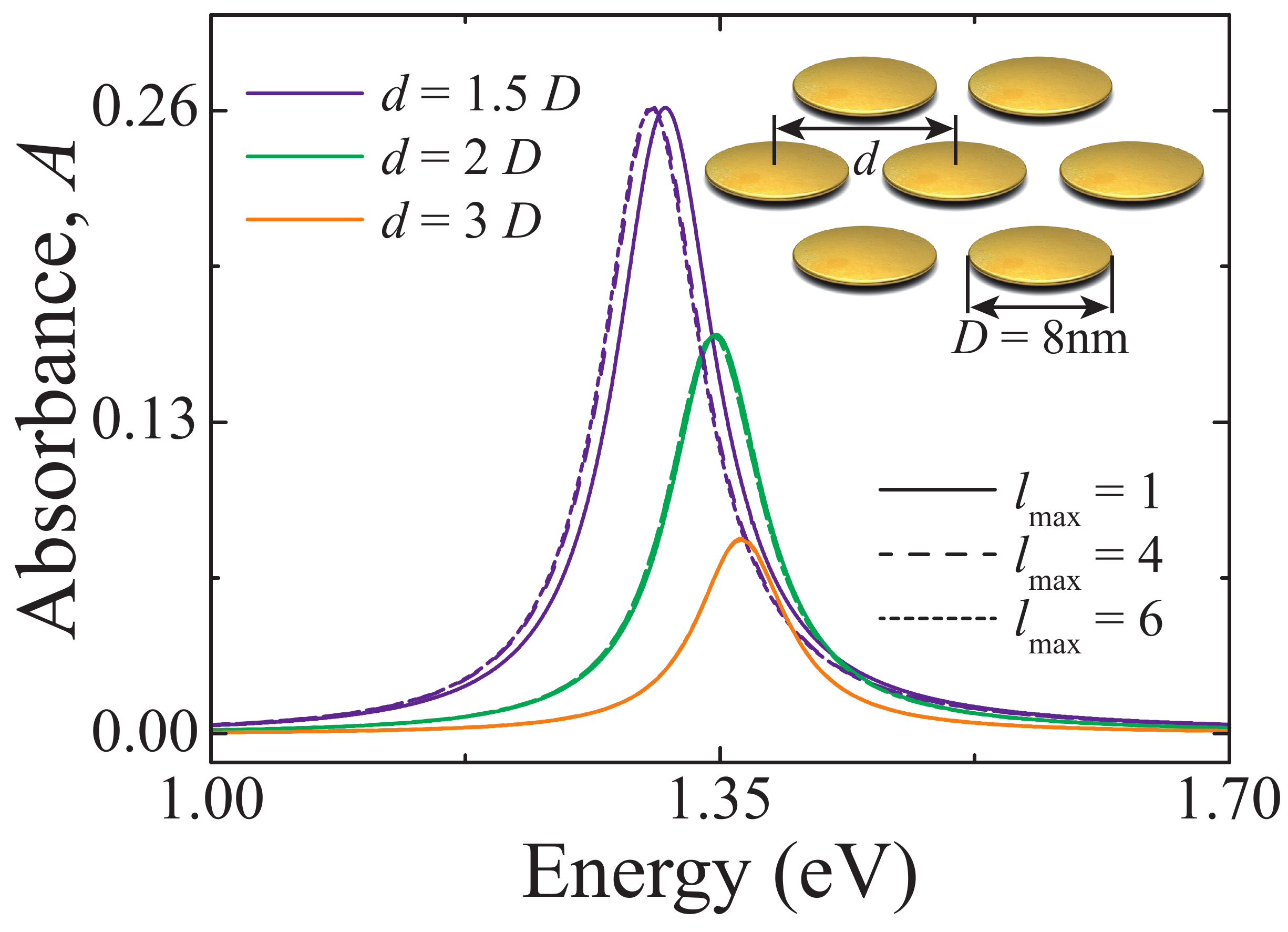}
\caption{{\bf Multipolar effects in the interaction of disk arrays.} We examine the validity of the dipole approximation to represent each disk in the arrays considered throughout this paper. For simplicity, we present classical results, as we expect that the conclusions should be directly applicable to quantum calculations as well. We use a layer KKR approach \cite{SYM98_1,SYM00} to calculate the absorbance of periodic hexagonal arrays of $D=8\,$nm gold disks with different lattice parameters $d$, with each disk represented through its scattering matrix, which is obtained with the boundary-element method (BEM). \cite{paper040} We include multiples of orbital angular momentum $l\le l_{\rm max}$. The results are remarkably converged already with $l_{\rm max}=1$ (dipolar approximation) for the two larger spacings under discussion, whereas for a period equal to 1.5 times the disk diameter multipolar corrections are rather small. Therefore, we conclude that the dipolar approximation used in this work provides quantitatively correct results for the geometrical parameters under consideration.
}\label{figS4}
\end{center}
\end{figure}

\begin{figure}
\begin{center}
\includegraphics[width=120mm,angle=0]{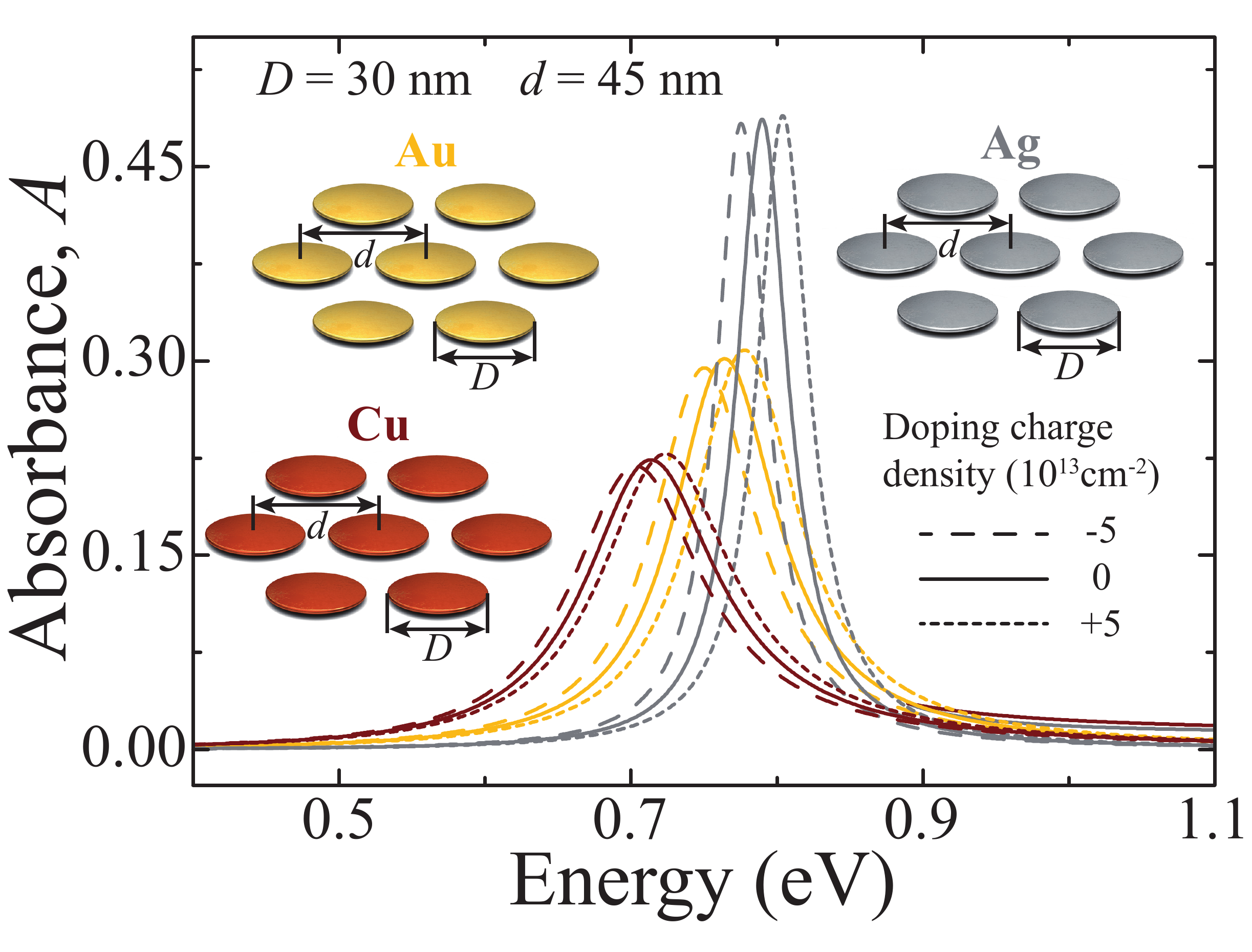}
\caption{{\bf Electrical modulation of the absorbance of hexagonal periodic arrays formed by single atomic-layer disks made of silver, gold, and copper.} The structures are similar to those of Fig.\ 5, but now the disks are larger, leading to lower-energy plasmons in the 0.7-0.8\,eV region. Silver is the less lossy of these three materials, and consequently, the optimum choice to maximize the optical tunability because its plasmons are narrower. Although the valence electron density is very similar in these metals, their plasmons occur at different energies due to variations in d-band screening. The geometrical and doping parameters are indicated by labels. The spectra are obtained from classical calculations.}\label{figS5}
\end{center}
\end{figure}

\begin{figure}
\begin{center}
\includegraphics[width=110mm,angle=0]{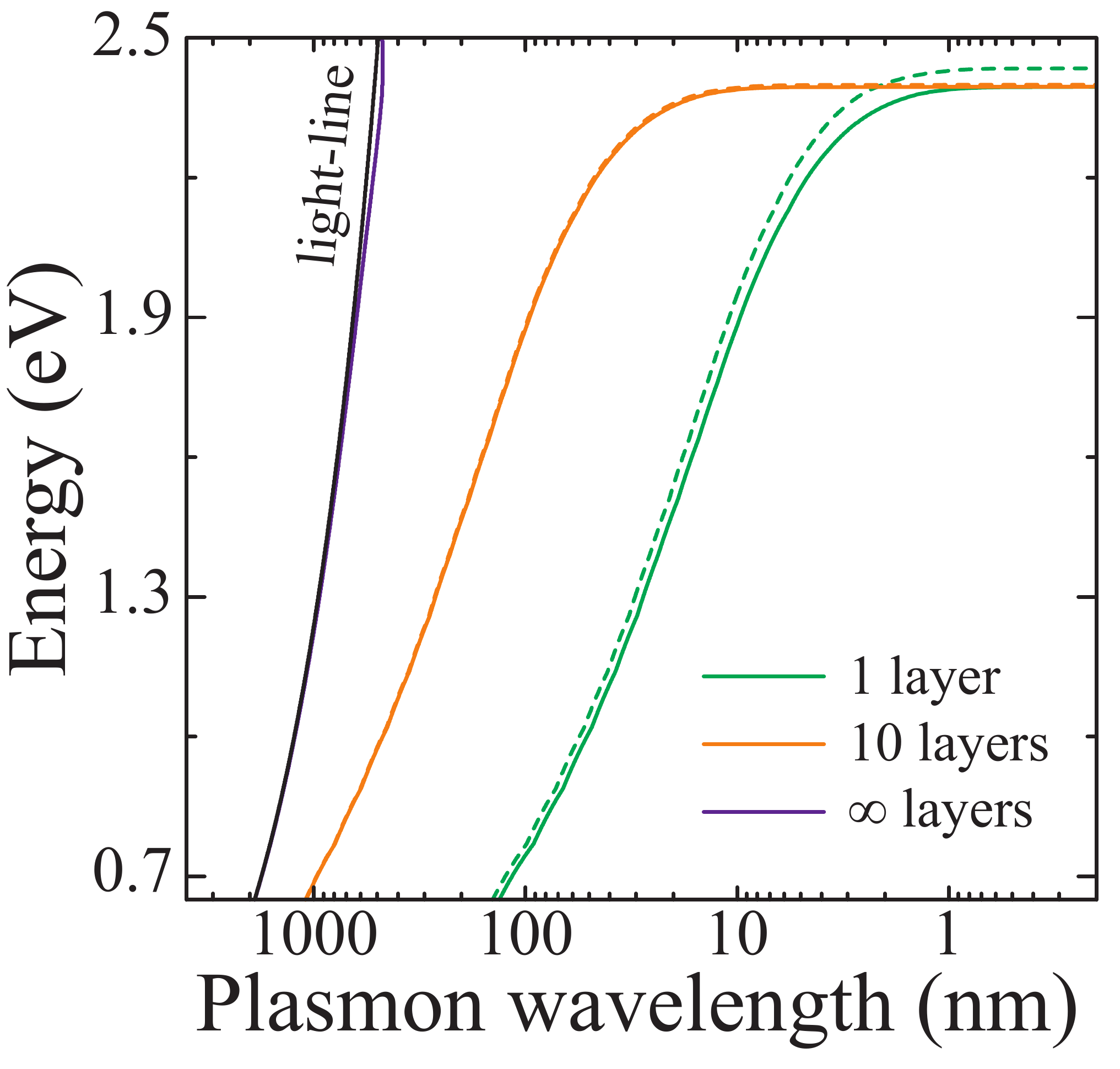}
\caption{{\bf Plasmons and tunability of thin homogeneous gold layers.} We represent the plasmon dispersion relations of layers consisting of 1 and 10 atomic monolayers oriented along the $(111)$ direction. The results are obtained from classical electromagnetic theory, using optical data for the dielectric function, \cite{JC1972} modified as described in the main text to include electrical doping. The plasmons of undoped layers (solid curves) are compared with those predicted for an additional density of $10^{14}\,$cm$^{-2}$ electrons (dashed curves). The light line and the dispersion of surface plasmons in a semi-infinite gold layer are shown for comparison. We conclude that the single-layer gold film can undergo similar electrical tunability as the nanostructures considered in the main text. The plasmons of the single layer are relatively far from the light line, although they have sizable wavelengths of 100's\,nm in the NIR spectral region. Like in graphene, the in/out-coupling of light to these plasmons represents a serious challenge, which can be overcome by decorating the layer with additional structures or by placing it near a grating of period comparable to the plasmon wavelength.}\label{figS6}
\end{center}
\end{figure}

\pagebreak

\section{Acknowledgments}

This work has been supported in part by the European Commission (Graphene Flagship CNECT-ICT-604391 and FP7-ICT-2013-613024-GRASP). A.M. acknowledges financial support from the Spanish MEC through the FPU program and from the Evans Attwell-Welch Postdoctoral Fellowship for Nanoscale Research, administered by the Richard E. Smalley Institute for Nanoscale Science and Technology.

%\bibliographystyle{achemso}
%\bibliography{../../bibtex/refs}

\providecommand*{\mcitethebibliography}{\thebibliography}
\csname @ifundefined\endcsname{endmcitethebibliography}
{\let\endmcitethebibliography\endthebibliography}{}

\end{document}